\documentclass[12pt]{article}

\usepackage{amssymb,amsmath}

\usepackage{graphicx}
\usepackage{color}
\usepackage[colorlinks=true
,urlcolor=blue
,citecolor=blue
,linkcolor=blue
,pagecolor=blue
,linktocpage=true
,pdfproducer=medialab
]{hyperref}
\usepackage[a4paper,width=15.5cm]{geometry}
\usepackage{subfigure}

\makeatletter \renewcommand{\@dotsep}{10000} \makeatother

\newcommand{\beq}{\begin{equation}}
\newcommand{\eeq}{\end{equation}}
\newcommand{\bea}{\begin{eqnarray}}
\newcommand{\eea}{\end{eqnarray}}

\begin{document}

\begin{center}

 {\Large

SO(10)  as a Framework for   Natural Supersymmetry

 } \vspace{1cm}

{\large   Ilia Gogoladze\footnote{E-mail: ilia@bartol.udel.edu\\
\hspace*{0.5cm} On  leave of absence from: Andronikashvili Institute
of Physics, 0177 Tbilisi, Georgia.},    \vspace{.3cm} Fariha Nasir\footnote {E-mail: fariha@udel.edu }  and  Qaisar Shafi\footnote{ E-mail:
shafi@bartol.udel.edu} } \vspace{.9cm}

{\baselineskip 20pt \it
Bartol Research Institute, Department of Physics and Astronomy, \\
University of Delaware, Newark, DE 19716, USA  } \vspace{.5cm}

\vspace{1.5cm}
 {\bf Abstract}
\end{center}

We consider an SO(10) grand unified theory in which the ratio of the $SU(2)_W$  and $SU(3)_c$  gaugino masses satisfy $M_2/M_3 \approx 3$, which results in the realization of natural supersymmetry.
 In the MSSM parameter space this relation looks artificial, but in the SO(10) case it results from a field with a designated  vacuum expectation value.   We consider two models, namely  $M_1:M_2:M_3=-1/5:3:1$ ($Case~I$), and $M_1:M_2:M_3=-5:3:1$ ($Case~II$).  Focusing on ameliorating the little hierarchy problem, we explore the parameter space of these models which yield small fine-tuning measuring parameters (natural supersymmetry) at the electroweak scale ($\Delta_{EW}$) as well as at the high scale ($\Delta_{HS}$). Although both models allow for the solution of the little hierarchy problem, the predicted sparticle spectra can differ markedly in the two cases. Depending on the ratio of the bino mass to the other gaugino masses, $Case~I$ leads to stau lepton masses of around a 100 GeV, while in $Case~II$, the stau slepton masses are in the several TeV range.
In $Case~I$, the bino-like neutralino can be as light as 90 GeV, while the gluino is heavier than 2 TeV or so. In $Case~II$, due to gluino-bino near degeneracy,  the bino cannot  be lighter than a TeV or so. Having a light neutralino with sizable bino-higgsino mixture in $Case~I$ allows the direct dark matter search experiments to test this class of models.

\newpage

\renewcommand{\thefootnote}{\arabic{footnote}}
\setcounter{footnote}{0}



\section{Introduction  \label{intro}}

The ATLAS and CMS experiments at the Large Hadron Collider (LHC)  have independently reported the discovery \cite{:2012gk, :2012gu} of a  Standard Model (SM) like Higgs boson resonance of mass $m_h \simeq 125-126$ GeV
 using the combined 7~TeV and 8~TeV data. This discovery is compatible with low scale supersymmetry,  since   the Minimal
Supersymmetric Standard Model (MSSM)  predicts an upper bound of $m_h \lesssim 135$ GeV for the lightest  CP-even Higgs boson  \cite{Carena:2002es}. On the other hand, no signals  for supersymmetry and the current lower bounds on the colored sparticle masses, namely
\begin{equation}
  m_{\tilde{g}} \gtrsim  1.4~{\rm TeV}~ ({\rm for}~ m_{\tilde{g}}\sim m_{\tilde{q}})~~~ {\rm and}~~~
m_{\tilde{g}}\gtrsim 0.9~{\rm TeV}~ ({\rm for}~ m_{\tilde{g}}\ll
m_{\tilde{q}}) ~\cite{Aad:2012fqa,Chatrchyan:2012jx})
\label{bound1}
\end{equation}
 have created some   skepticism about naturalness arguments for  low scale supersymmetry.
Although the sparticle mass bounds in Eq. (\ref{bound1}) are mostly derived for R-parity conserving constrained MSSM (CMSSM), they are  more or less  applicable for a significant class of low scale supersymmetric models.  It  was discussed in ref. \cite{Bhattacherjee:2013gr} that there is room in the MSSM parameter space where
the bounds in Eq. (\ref{bound1}) can be relaxed, but this parameter space is not large  and the models are  specific.
Low scale supersymmetry can accommodate a Higgs with mass $m_h \simeq 125 \rm \ GeV$ in the MSSM but requires either a very large, ${\cal O} (\mathrm{few}-10)$ TeV,  stop quark mass \cite{Ajaib:2012vc}, or a large soft supersymmetry  breaking (SSB) trilinear $A_t$-term, with a stop quark mass of around a TeV \cite{Djouadi:2005gj}.
 A heavy stop quark can lead to the   ``little hierarchy"   problem \cite{b5}.

 In the MSSM, through minimizing the tree level scalar potential,
the $Z$ boson mass, $M_Z=91.2$ GeV,   can be computed in terms of the supersymmetric bilinear Higgs parameter ($\mu$) and the SSB $mass^2$  terms for the up ($m_{H_{u}}$) and down ($m_{H_{d}}$)-type Higgs doublets \cite{mssm}
\begin{equation}
\frac{1}{2}M_{Z}^{2}= -\mu
^{2}+\left(\frac{m_{H_{d}}^{2} - m_{H_{u}}^{2}  \text{
tan}^{2}\beta}{\text{ tan}^{2}\beta -1} \right) \approx -\mu
^{2}-m_{H_{u}}^{2}. \label{e4}
\end{equation}
The approximation in Eq. (\ref{e4}) works well for moderate and large $\tan\beta$ values.
We see from Eq. (\ref{e4}) that unless $\mu^2$ and $m^{2}_{H_{u}}$  values   are of order $M^{2}_Z$,  some fine-tuning of the two parameters is required, which is an indication of the so called   ``little hierarchy" problem.  In order to clarify the fine tuning condition among  SSB parameters  and $\mu$ term in the MSSM we need to check the relation between these parameters not only at the  electroweak ($M_{EW}$) scale but at any scale up to the
grand unified theory (GUT) or Planck  scale  \cite{Baer:2012mv,Gogoladze:2012yf}. In this paper we pursue this approach by seeking the parameter space which yields $\mu$ and $m_{H_{u}}$ values comparable to the $Z$-boson mass at any scale, with the squarks and gluinos being much heavier than a TeV at $M_{EW}$.

It has been shown in
\cite{Gogoladze:2012yf,Abe:2007kf,Younkin:2012ui}  and references therein that non-universal gaugino masses at  GUT scale ($M_{\rm GUT}$)
 can help resolve the little hierarchy problem. In Particular, it was shown  in ref. \cite{Gogoladze:2012yf,Abe:2007kf,Younkin:2012ui}
 that the little hierarchy problem  can be largely resolved if the ratio between $SU(2)_L$ and $SU(3)_c$
gaugino masses satisfy the asymptotic relation $M_2/M_3\approx 3$.
In this case the leading contributions to  $m^{2}_{H_{u}}$ through  RGE  evolution  are proportional to $M_2$ and $M_3$ can cancel each other. This allows for large  values of  $M_2$ and $M_3$ in the
gravity mediated supersymmetry  breaking scenario \cite{Chamseddine:1982jx}, while keeping the value of $m^{2}_{H_{u}}$ relatively small. On the other hand, large values of $M_2$ and $M_3$ yield a heavy stop quark ($>$ TeV) which is necessary in order to accommodate   $m_h \simeq 125$ GeV.
Note that  except for the bino, all remaining sparticle masses weakly depend on the   $U(1)_Y$ gaugino mass ($M_1$), especially if we require a neutralino LSP. This implies that the ratio of $M_1$ to other MSSM gauginos is not important for the little hierarchy problem, but is very important for the neutralino to be a suitable dark matter candidate.  This was one of the main  motivations to study the
  $SU(4)_c \times SU(2)_L \times SU(2)_R$ (4-2-2) model \cite{Pati:1974yy}.
In the 4-2-2 model with C-parity  \cite{c-parity}, the number of  independent gaugino masses reduces from  three  to two
 and for bino mass we have:  $M_1 = {\frac{2}{5}}M_3+{\frac{3}{5}}M_2$. 
 
 In our previous paper \cite{Gogoladze:2012yf},  employing the ISAJET 7.84 package \cite{ISAJET}, we showed that  in the 4-2-2 model there exists region of SSB parameter space in which the little hierarchy problem can be ameliorated  not only at the $M_{\rm EW}$ but also at $M_{\rm GUT}$.   The corresponding solutions  satisfy all collider constraints with the LSP
 neutralino being a suitable dark matter candidate.
Since  ameliorating the little hierarchy  problem in the non-universal gaugino case occurs only for $M_2/M_3\approx 3$  \cite{Gogoladze:2012yf},  it is hard to explain the origin of this ratio within the 4-2-2 model or in the MSSM. Therefore, in this paper we study specific GUT scenarios  where the desired  ratio for the MSSM gaugino masses can be realized.
There are several ways to obtain $M_2/M_3\approx 3$ at the GUT scale, for instance, by using string or GUT contractions \cite{Younkin:2012ui}. On the other hand, the low energy sparticle spectrum strongly depends on the value of $M_1$ and as we will show, varying the ratios  $M_2/M_1$ and  $M_3/M_1$ yields very distinguishable low energy sparticle spectrum which can be tested at LHC.

In this paper we consider two cases realized in the content of SO(10) : $M_1:M_2:M_3=-1/5:3:1$ ($Case~I$), and $M_1:M_2:M_3=-5:3:1$ ($Case~II$). These asymptotic relation among gauginos can arise
by considering suitable supersymmetry breaking F-term contributions to the MSSM gaugino masses.
In this article  we study  $Case~I$ and $Case~II$  separately  since they yield very specific low energy spectrum which can be tested at the LHC. It is interesting to note that in $Case~II$ we obtain approximate unification for MSSM gaugino masses at low scale
which can lead to the so-called `compressed' supersymmetry \cite{Martin:2007gf}.

The layout of this paper is as follows.   In Section \ref{so10-model}, we describe how the MSSM gaugino mass relations can be obtained at $M_{\rm GUT}$. We present in Section \ref{so10-model}   the parameter space that we scan over and the experimental constraints that  we employ.
In Section \ref{ft-11} we briefly describe the fine-tuning conditions at low and high scales.
The  results for $Case~I$ and $Case~II$  are discussed in  Section \ref{results-99}.
Our conclusion are presented in Section \ref{conclusions}.

\section{ Parameters and  Phenomenological Constraints \label{so10-model}}

 It has been pointed out \cite{Martin:2009ad} that non-universal MSSM gaugino masses at $ M_{\rm GUT} $ can arise from non-singlet $F$-terms, compatible with the underlying GUT symmetry  such as SU(5) or SO(10). The SSB
gaugino masses in supergravity  \cite{Chamseddine:1982jx} can arise  from the following
non renormalizable  operator:
\begin{align}
 -\frac{F^{ab}}{2 M_{\rm
P}} \lambda^a \lambda^b + {\rm c.c.}
\end{align}
 Here $\lambda^a$ is the two-component gaugino field, $ F^{ab} $ denotes the $F$-component of the field which breaks SUSY, and the indices $a,b$ run over
the adjoint representation {of the gauge group}. The resulting gaugino
mass matrix is $\langle F^{ab} \rangle/M_{\rm P}$, where the
supersymmetry breaking  parameter $\langle F^{ab} \rangle$
transforms as a singlet under the MSSM gauge group $SU(3)_{c}
\times SU(2)_L \times U(1)_Y$. The $F^{ab}$ fields belong to an
irreducible representation in the symmetric part of the direct product of the
adjoint representation of the unified group. This is a supersymmetric generalization of operators considered a long time ago \cite{Hill:1983xh}.

In SO(10), for example,
\begin{align}
({ 45} \times { 45} )_S = { 1} + { 54} + { 210} +
{ 770}
\end{align}
If the  $F$-component of the field transforms as a  210 or 770 dimensional
representation of SO(10) \cite{Martin:2009ad}, one  obtains the following relation
among the MSSM gaugino masses at $ M_{\rm GUT}$, depending whether we  have a ``normal" embedding
\cite{SO10GUT} of $SU(5)\times U(1)\subset SO(10)$
\begin{align}
~~~~~~~~~~~~~~~~~~~~~~~~ M_1: M_2:M_3= -1/5:3:1, ~~~~~  ~~~~~~~~~~~~~~~~~ Case~I
\label{gaugino10-1}
\end{align}
or ``flipped" $SU(5)$ embedding \cite{flipped},  $SU(5)^{\prime} \times U(1)\subset SO(10)$:
\begin{align}
~~~~~~~~~~~~~~~~~~~~~~ M_1: M_2:M_3= -5:3:1. ~~~~~~~ ~~~~~~~~~~~~~~~~~ Case~II
\label{gaugino10-2}
\end{align}
Here $M_1, M_2, M_3$ denote the  gaugino masses of $U(1)$, $SU(2)_L$ and $SU(3)_c$
respectively.

Notice that in  general, if $F^{ab}$ transforms non trivially under SO(10), the SSB terms such as the trilinear couplings and scalar mass terms are not necessarily universal at $M_{\rm GUT}$. However, we can assume, consistent with SO(10) gauge symmetry, that the coefficients associated with terms that violate SO(10) invariance are suitably small, except for the gaugino term in Eq.(\ref{gaugino10-1}) (or in Eq.(\ref{gaugino10-2})). We also assume that D-term contributions to the SSB terms are much smaller compared to the  contributions from fields with non-zero auxiliary F-terms.

Employing the boundary condition from Eq.(\ref{gaugino10-1}), one  can define the MSSM gaugino masses at $ M_{\rm GUT} $ in terms of the mass parameter $M_{1/2}$ :
\begin{align}
M_1= -1/5 M_{1/2},~~  M_2= 3M_{1/2},~~ M_3=  M_{1/2}, ~~~~~~~~~~~~~~~ Case~ I
 \label{case1}
\end{align}
 and
\begin{align}
M_1= -5 M_{1/2},~~  M_2= 3 M_{1/2},~~ M_3=  M_{1/2}. ~~~~~~~~~~~~~~~ Case~ II
 \label{case1}
\end{align}

Thus, we have the following fundamental parameters in the SSB sector:
\begin{align}
m_{0},~~ M_{1/2},~~ A_0,~~ \tan\beta.
\label{params}
\end{align}
 In this paper we  focus on $\mu>0$.

We have performed random scans with the ISAJET~7.84 package~\cite{ISAJET} for the following range of the parameter space:
\begin{align}
0\leq  m_{0}=m_{H_{u}}=m_{H_{d}}  \leq 20\, \rm{TeV} \nonumber \\
0 \leq M_{1/2}  \leq 5 \, \rm{TeV} \nonumber \\
2\leq \tan\beta \leq 60 \nonumber \\
-3\leq A_{0}/m_{0} \leq 3\nonumber\\
\mu > 0
 \label{parameterRange}
\end{align}
We set    $m_t = 173.2\, {\rm GeV}$  \cite{Aaltonen:2012ra}
and  $m_b(m_Z)=2.83$ GeV, which is hard-coded into ISAJET. Note that  varying the top quark mass within a $1\,\sigma$ interval can change  the Higgs boson mass by 1 GeV or so \cite{Gogoladze:2012ii}.

In performing the random scan a uniform and logarithmic distribution of random points is first generated in the parameter space given in Eq. (\ref{parameterRange}).
The function RNORMX \cite{Leva} is then employed
to generate a Gaussian distribution around each point in the parameter space.
The collected data points all satisfy
the requirement of REWSB with one of the neutralinos being the lightest supersymmetric particle (LSP).

After collecting the data, we impose
the mass bounds on all the particles \cite{Nakamura:2010zzi} and use the
IsaTools package~\cite{Baer:2002fv}
to implement the various phenomenological constraints. We successively apply the following experimental constraints on the data that
we acquire from ISAJET:
\begin{table}[h]\centering
\begin{tabular}{rlc}
$1.7 \times 10^{-9} \leq\, BR(B_s \rightarrow \mu^+ \mu^-) $&$ \leq\, 4.7 \times 10^{-9}$        & \cite{Aaij:2012hcp}     \\
$2.85 \times 10^{-4} \leq BR(b \rightarrow s \gamma) $&$ \leq\, 4.24 \times 10^{-4} \;
 (2\sigma)$ &   \cite{Barberio:2008fa}  \\
$0.15 \leq \frac{BR(B_u\rightarrow
\tau \nu_{\tau})_{\rm MSSM}}{BR(B_u\rightarrow \tau \nu_{\tau})_{\rm SM}}$&$ \leq\, 2.41 \;
(3\sigma)$ &   \cite{Barberio:2008fa}  \\
 $ 0 \leq \Delta(g-2)_{\mu}/2 $ & $ \leq 55.6 \times 10^{-10} $ & \cite{Bennett:2006fi}
\end{tabular}\label{table}
\end{table}

\hspace{-10mm} Note that for $\Delta(g-2)_{\mu}$, we only require that the  model does no worse than the SM.

\section{Fine-Tuning Constraints \label{ft-11}}
The latest (7.84) version of  ISAJET \cite{ISAJET} calculates the  fine-tuning conditions related to the little hierarchy problem at $M_{EW}$
and at the GUT scale ($M_{HS}$). We will briefly describe these parameters in this section.

After including the one-loop effective potential contributions to the tree level MSSM Higgs potential, the Z boson mass is given by the  relation:
\begin{equation}
\frac{M_Z^2}{2} =
\frac{(m_{H_d}^2+\Sigma_d^d)-(m_{H_u}^2+\Sigma_u^u)\tan^2\beta}{\tan^2\beta
-1} -\mu^2 \; .
\label{eq:mssmmu}
\end{equation}
The $\Sigma$'s stand for the contributions coming from the one-loop effective potential (For more details see ref. \cite{Baer:2012mv}). All parameters  in Eq. (\ref{eq:mssmmu}) are defined at the weak scale $M_{EW}$.

\subsection{Electroweak Scale Fine-Tuning}
\label{esft}

In order to measure the EW scale fine-tuning condition associated with the little hierarchy problem, the following definitions are used \cite{Baer:2012mv}:
\begin{equation}
 C_{H_d}\equiv |m_{H_d}^2/(\tan^2\beta -1)|,\,\, C_{H_u}\equiv
|-m_{H_u}^2\tan^2\beta /(\tan^2\beta -1)|, \, \, C_\mu\equiv |-\mu^2 |,
\label{cc1}
\end{equation}
 with
each $C_{\Sigma_{u,d}^{u,d} (i)}$  less
than some characteristic value of order $M_Z^2$.
Here, $i$ labels the SM and supersymmetric
particles that contribute to the one-loop Higgs potential.
For the fine-tuning condition we have
\begin{equation}
 \Delta_{\rm EW}\equiv {\rm max}(C_i )/(M_Z^2/2).
\label{eq:ewft}
\end{equation}
Note that Eq. (\ref{eq:ewft}) defines the fine-tuning  condition  at $M_{EW}$ without addressing
the question of the origin of the parameters that are involved.

\subsection{High Scale Fine-Tuning}
\label{hsft}

In most SUSY breaking scenarios the parameters in
Eq.~(\ref{eq:mssmmu}) are defined at a scale higher than $M_{EW}$.
In order to fully address  the fine-tuning condition we need to  check the relations
 among the parameters involved  in Eq.~(\ref{eq:mssmmu}) at high scale. We relate the parameters at low and high scales  following ref. \cite{Baer:2012mv}:
 \begin{equation}
 m_{H_{u,d}}^2=
m_{H_{u,d}}^2(\Lambda) +\delta m_{H_{u,d}}^2, \ \,\,\,
\mu^2=\mu^2(\Lambda)+\delta\mu^2.
\end{equation}
 Here
$m_{H_{u,d}}^2(\Lambda)$ and $\mu^2(\Lambda)$ are the corresponding
parameters renormalized at the high scale $\Lambda$, which can either be the GUT, string or some other scale.
The $\delta m_{H_{u,d}}^2$, $\delta\mu^2$ measure how the given parameter is changed due to Renormalization Group Evolution (RGE).
 Eq.~(\ref{eq:mssmmu}) can be re-expressed in the form
\begin{eqnarray}
\frac{m_Z^2}{2} = \frac{(m_{H_d}^2(\Lambda)+ \delta m_{H_d}^2 +
\Sigma_d^d)-
(m_{H_u}^2(\Lambda)+\delta m_{H_u}^2+\Sigma_u^u)\tan^2\beta}{\tan^2\beta -1}
- (\mu^2(\Lambda)+\delta\mu^2).
\label{eq:FT}
\end{eqnarray}
Following ref. \cite{Baer:2012mv} we define:
\begin{eqnarray}
&B_{H_d}\equiv|m_{H_d}^2(M_{HS})/(\tan^2\beta -1)|,
B_{\delta H_d}\equiv |\delta m_{H_d}^2/(\tan^2\beta -1)|, \nonumber \\
&B_{H_u}\equiv|-m_{H_u}^2(M_{HS})\tan^2\beta /(\tan^2\beta -1)|, B_{\mu}\equiv|\mu^2(M_{HS})|, \nonumber \\
&B_{\delta H_u}\equiv|-\delta m_{H_u}^2\tan^2\beta /(\tan^2\beta -1)|,
  B_{\delta \mu}\equiv |\delta \mu^2|,
  \label{bb1}
\end{eqnarray}
and  the
high scale fine-tuning measure $\Delta_{\rm HS}$ is defined to be
\begin{equation}
\Delta_{\rm HS}\equiv {\rm max}(B_i )/(M_Z^2/2).
\label{eq:hsft}
\end{equation}

The current experimental bound on the chargino mass ($m_{\widetilde W}> 103$ GeV) \cite{Nakamura:2010zzi} indicates that either $\Delta_{EW}$ or $\Delta_{HS}$ cannot be less than 1. The quantities $\Delta_{EW}$ and  $\Delta_{HS}$ measure the sensitivity of the Z-boson mass to the parameters defined in Eqs. (\ref{cc1}) and (\ref{bb1}), such that $(100/\Delta_{EW})\%$  ($(100/\Delta_{HS})\%$) is the degree of fine-tuning at the corresponding scale.


\section{Results \label{results-99}}



\begin{figure}[th!]
\begin{center}
\includegraphics[width=7.6cm,height=6.cm]{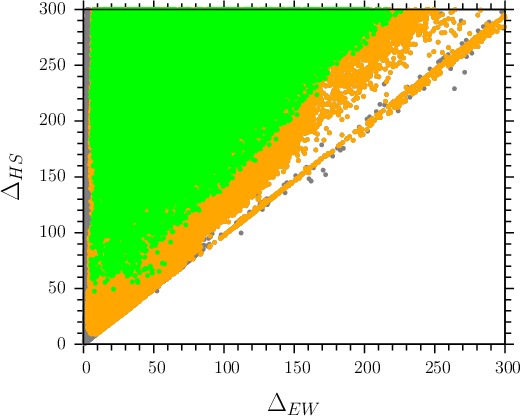}
\end{center}
\caption{Plots in  the $\Delta_{HS}- \Delta_{EW}$ planes for the $case~I$. Gray points are consistent with REWSB and neutralino to be the LSP. Orange points form a subset of the
gray ones and satisfy all the constraints described in Section \ref{so10-model}.  Green points belong to the subset of orange points
and satisfy the Higgs mass range $123\, {\rm GeV} \leq m_h \leq 127 \,{\rm GeV}$.}
\label{fig-c1}
\end{figure}


\begin{figure}[]
\begin{center}
\includegraphics[width=7.2cm,height=6.cm]{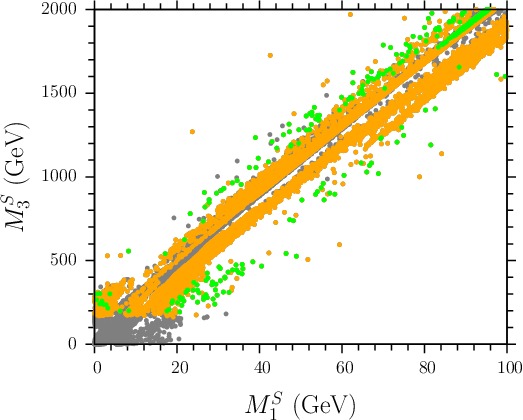}
\includegraphics[width=7.2cm,height=6.cm]{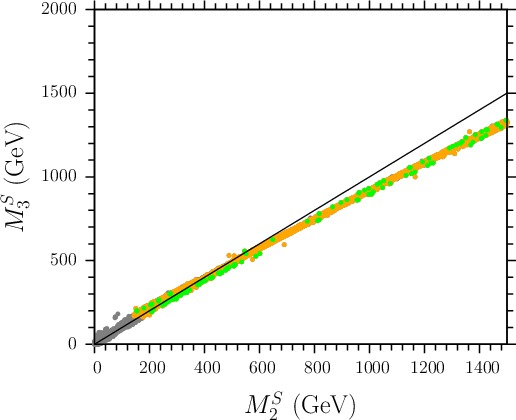}
\end{center}
\caption{Plots in $M^S_3 - M^S_1$ and $M^S_3-M^S_2$ planes for the $Case~I$. Color coding
is the same as described in Figure \ref{fig-c1}.}
\label{fig-c2}
\end{figure}


\begin{figure}[]
\begin{center}
\includegraphics[width=7.2cm,height=6.cm]{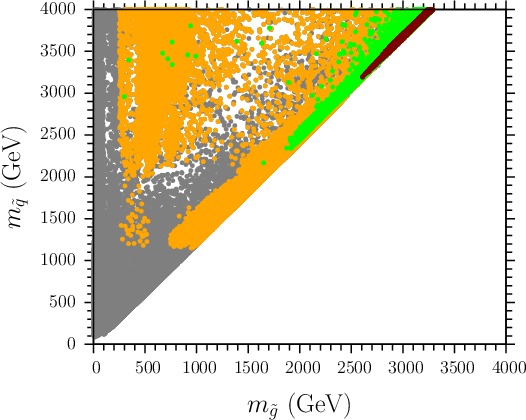}
\includegraphics[width=7.2cm,height=6.cm]{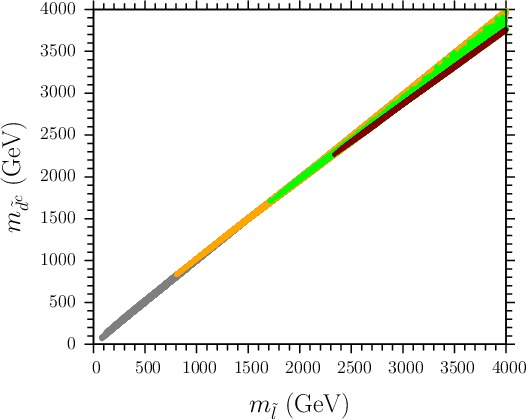}
\end{center}
\caption{Plots in $m_{\tilde{q}} - m_{\tilde{g}}$ and $m_{\tilde{d}^c} - m_{\tilde{l}}$  planes for the $Case~I$. Color coding
is the same as described in Figure \ref{fig-c1}.  In addition, we have used maroon color to denote a subset of the green points with $\Delta_{HS}<100$ and $\Delta_{EW}<100$.  }
\label{fig-c3}
\end{figure}


\begin{figure}[]
\begin{center}

\includegraphics[width=7.2cm,height=6.cm]{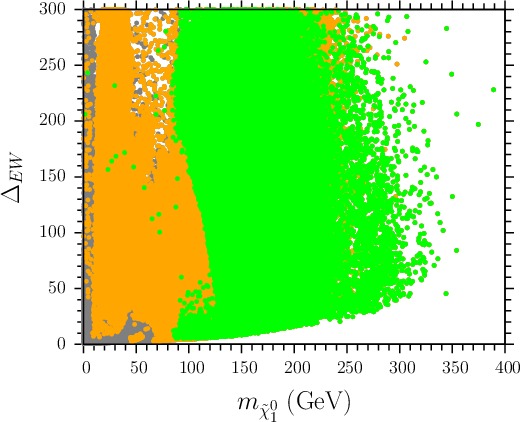}
\includegraphics[width=7.2cm,height=6.cm]{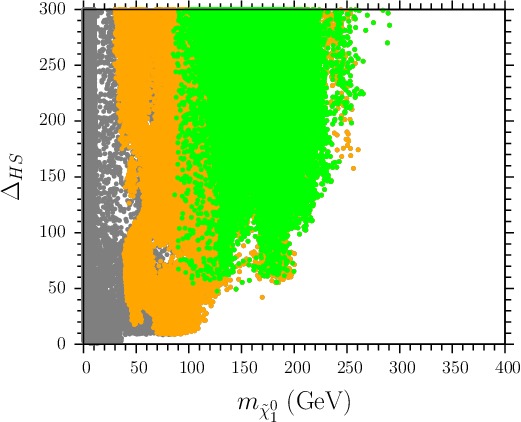}\vspace*{3mm}
\includegraphics[width=7.2cm,height=6.cm]{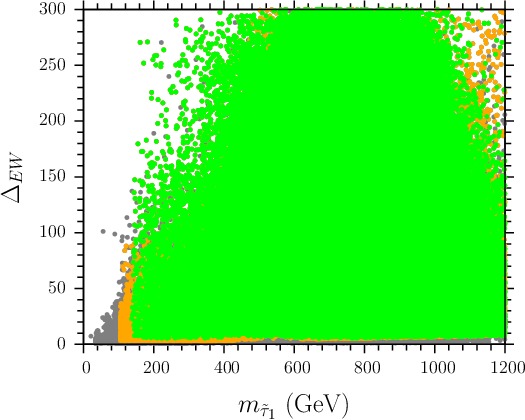}
\includegraphics[width=7.2cm,height=6.cm]{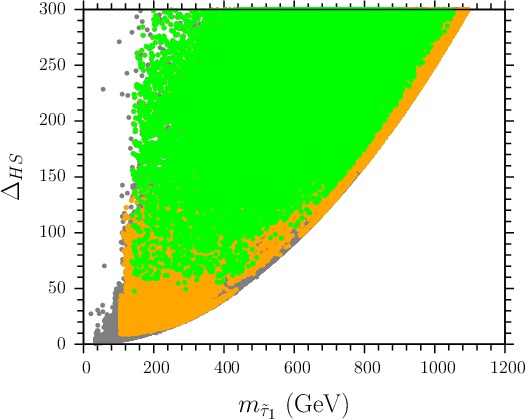}\vspace*{3mm}
\includegraphics[width=7.2cm,height=6.cm]{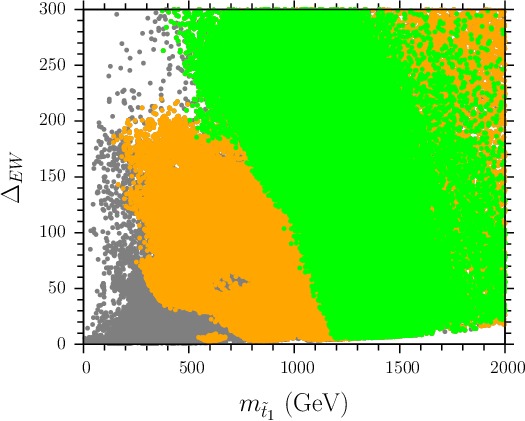}
\includegraphics[width=7.2cm,height=6.cm]{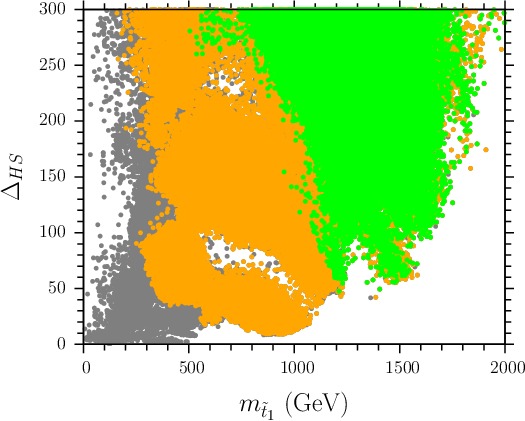}
\end{center}
\caption{Plots in
 $\Delta_{EW}-m_{\tilde{\chi}^{0}_1}$,  $\Delta_{HS}-m_{\tilde{\chi}^{0}_1}$,
 $\Delta_{EW}-m_{\tilde{\tau}_1}$,  $\Delta_{HS}-m_{\tilde{\tau}_1}$,
 $\Delta_{EW}-m_{\tilde{t}_1}$,  $\Delta_{HS}-m_{\tilde{t}_1}$,
 planes for the $Case~I$ cases. Color coding
is the same as described in Figure \ref{fig-c1}.   }
\label{fig-c4}
\end{figure}


\begin{figure}[t!]
\begin{center}
\includegraphics[width=7.2cm,height=6.cm]{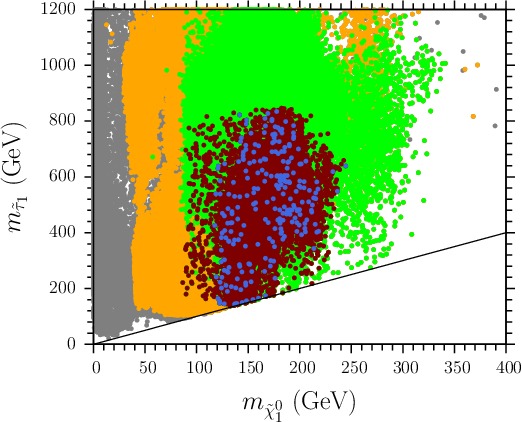}\vspace*{3mm}
\includegraphics[width=7.2cm,height=5.6cm]{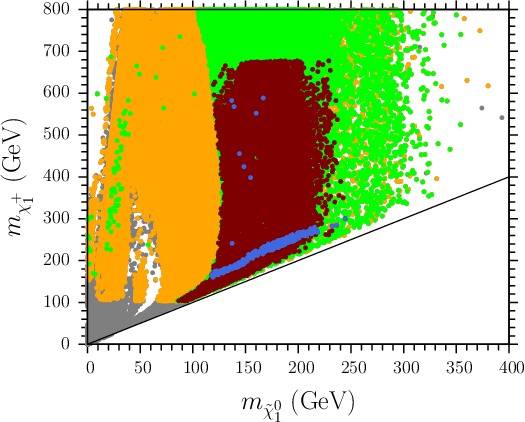}
\end{center}
\caption{Plots in  the
 $m_{\tilde{\tau}_1}-m_{\tilde{\chi}^{0}_1}$ and $m_{\tilde{\chi}^{+}_1}-m_{\tilde{\chi}^{0}_1}$ planes for the $Case~I$. Color coding
is the same as described in Figure \ref{fig-c3}. In addition, we have used blue  color to denote a subset of the maroon points with $\Omega h^2<1$. The unit slope line is shown to guide the eye.  }
\label{fig-c4}
\end{figure}


\begin{figure}[]
\begin{center}
\includegraphics[width=7.2cm,height=6.cm]{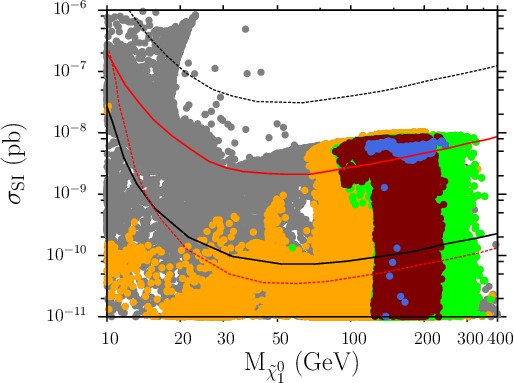}
\includegraphics[width=7.2cm,height=6.cm]{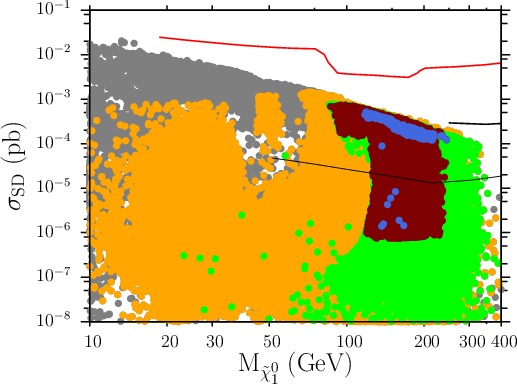}
\end{center}
\caption{Plots in $\sigma_{SI}- m_{\tilde{\chi}^{0}_1}$ and  $\sigma_{SD}- m_{\tilde{\chi}^{0}_1}$ planes for the $Case~I$. Color coding
is the same as described in Figure \ref{fig-c4}. The left panel shows the current and future bounds from CDMS as black (solid and dashed) lines, and as red (solid and dotted) lines for the Xenon experiment. The right panel also shows the current bounds from Super
K (solid red line) and IceCube (solid black line), and future reach of
IceCube DeepCore (dotted black line).  }
\label{fig-c5}
\end{figure}


\subsection{$Case~I$ \label{model-I}}

Figure \ref{fig-c1} shows our results in the $\Delta_{HS}-\Delta_{EW}$ planes for $Case~I$ where we assume the following relation
among the MSSM gauginos: $M_1: M_2:M_3= -1/5:3:1$. Gray points are consistent with REWSB and a neutralino LSP. The orange points form a subset of the
gray ones and satisfy all the constraints described in Section \ref{so10-model}.  Green points form a subset of orange points
and satisfy the condition $123\, {\rm GeV} \leq m_h \leq 127 \,{\rm GeV}$ imposed on higgs mass.
The green points in the $\Delta_{HS}-\Delta_{EW}$ plane indicate that it is possible to simultaneously have $\Delta_{EW}\approx 10$ and
$\Delta_{HS}\approx 40$. This result is somewhat similar to what was obtained in the 4-2-2 framework \cite{Gogoladze:2012yf}, but there is significant difference in terms of neutralino as the possible dark matter candidate.  It was shown that in the 4-2-2 model with C-parity, due to the relation $M_3/M_1>2/5$ between bino and gluino masses, the lightest neutralino is mostly Higgsino, or when the neutralino is mostly bino its mass is more then a few hundred GeV. On the other hand in $Case~I$ we have $M_3/M_1=5$ at the GUT scale,  and we can observe  in the $M_3^S-M_1^S$ plane of Figure \ref{fig-c2} that the 
bino-like neutralino can be as light as 30 GeV, with the gluino heavier than 900 GeV and the lightest CP even Higgs around 125 GeV.
This means that we can still have the light Higgs (h) and Z resonance channels for the neutralino dark matter candidate. Note that it  is difficult to have such a light neutralino in the case of gaugino unification at the GUT scale.
 In Figure \ref{fig-c2},  $M_1^S$, $M_2^S$ and  $M_3^S$  stand for $U(1)_Y$, $SU(2)_W$ and $SU(3)_c$ gaugino mass at the SUSY breaking scale. Color coding is the same as described  in Figure \ref{fig-c1}.

 In the $M_3^S-M_2^S$ plane of Figure \ref{fig-c2}, we can see that the gluinos are somewhat lighter than the winos, which results in mass degeneracy of the first two families of right handed squarks and left handed sleptons as seen in Figure \ref{fig-c3} . More interesting perhaps is the parameter space shown in Figure \ref{fig-c3} where squarks are shown to be lighter than the sleptons.
 Color coding is the same as described  in Figure \ref{fig-c1}. In addition, we have used the maroon color to denote a subset of the green points with $\Delta_{HS}<150$ and $\Delta_{EW}<100$. From the $m_{\tilde{d^c}}-m_{\tilde{l}}$ plane, it is interesting to note that some of these green points correspond to lighter
 right handed squark, left handed sleptons and gluinos compared to the solution with smaller  $\Delta_{HS}$ and $\Delta_{EW}$ presented in maroon color.

 The solution to the little hierarchy problem also prefers heavier gluinos as seen from the  $m_{\tilde{q}}-m_{\tilde{g}}$ plane of Figure \ref{fig-c3}. Here
$m_{\tilde{q}}$ stands for first two generation of left handed squarks. Our conclusion is that in $Case~I$
the solution which  ameliorates the little hierarchy problem yields  gluino mass  heavier than 2.5 TeV  and is therefore fully consistent with the current CMS and ATLAS observations \cite{Aad:2012fqa,Chatrchyan:2012jx}.

In order to gain a sense of the sparticle mass ranges expected from natural supersymmetry considerations we plot $\Delta_{EW}$ and $\Delta_{HS}$ versus selected sparticle masses. We independently present $\Delta_{EW}$ and $\Delta_{HS}$ dependence on the sparticle masses since there are different approaches to the little hierarchy problem. For instance, in the framework of the so-called radiative natural SUSY \cite{Baer:2012cf}, radiatively-induced low scale-tuning  is addressed. If one adopts this approach we can only focus on  $\Delta_{EW}$ versus sparticle masses. For a more general case we can check how the little hierarchy condition works for any scale presenting  results in terms of  $\Delta_{EW}$ and $\Delta_{HS}$. In Figure  \ref{fig-c4} the color coding is the same as described  in Figure \ref{fig-c1}.
The $\Delta_{EW}-m_{\tilde{\chi}^0_1}$ plane shows that in $Case~I$ the LSP neutralino can be as light as 80 GeV with $\Delta_{EW}\thickapprox 10$. However, the $\Delta_{HS}-m_{\tilde{\chi}^0_1}$ plane shows that $\Delta_{HS}\thickapprox 60$ for the same value of  neutralino mass.

In the $\Delta_{EW}-m_{\tilde{\tau}_1}$ plane, we show that $\Delta_{EW}\thickapprox 10$ allows $m_{\tilde{\tau}_1}\thicksim 100-1000$ GeV. On the other hand  $\Delta_{EW}< 100$ allows $m_{\tilde{\tau}_1}\thicksim 100-600$ GeV, which is accessible at the ILC. According to the $\Delta_{EW}-m_{\tilde{\tau}_1}$ and $\Delta_{HS}-m_{\tilde{t}_1}$ plots the stop quarks are relatively heavy compared to the stau leptons.
 The  $\Delta_{EW}-m_{\tilde{t}_1}$ plane shows that the lightest top in $Case~I$ has mass  $m_{\tilde{t}_1}\thickapprox 1200$ GeV with   $\Delta_{EW}\thickapprox 10$ .

We can see in Figure \ref{fig-c4} that a variety of coannihilation
 scenarios are compatible with the relaxed little hierarchy problem
and neutralino dark matter. Color coding
is the same as described in Figure \ref{fig-c3}. In addition, we have used blue  color to denote a subset of the maroon points with $\Omega h^2<1$. The unit slope line  indicates the
presence of neutralino-stau coannihilation scenario and bino-higgsino admixture for the lightest
neutralino. In the $m_{\tilde{\tau}_1}-m_{\tilde{\chi}^{0}_1}$ plane of Figure \ref{fig-c4} we see that neutralino-stau coannihilation scenario with low fine tuning condition is realized for $m_{\tilde{\tau}_1} \approx 150$ GeV, which can be very easily tested at the ILC.
Bino-higgsino dark matter can be realized for   $120 \lesssim m_{\tilde{\chi}^{0}_1} \lesssim 230$ GeV, as seen in the $m_{\tilde{\chi}^{+}_1}-m_{\tilde{\chi}^{0}_1}$ panel.

In Figure \ref{fig-c5} we show the implication of our analysis for direct detection of dark matter. Plots are shown in the  $\sigma_{\rm SI}  -  m_{\tilde{\chi}_1^{0}}$ and $\sigma_{\rm SD}  -  m_{\tilde{\chi}_1^{0}}$
planes and the color coding
is the same as described in Figure \ref{fig-c4}. The left panel shows the current and future bounds from CDMS as black (solid and dashed) lines, and as red (solid and dotted) lines for the Xenon experiment. The right panel also shows the current bounds from Super
K (solid red line) and IceCube (solid black line), and future reach of
IceCube DeepCore (dotted black line).
We found that the  $\mu \sim
M_1$  case is consistent with natural supersymmetry, as this is the
requirement to get bino-higgsino admixture for the lightest
neutralino which, in turn, enhances both the spin dependent and spin
independent neutralino-nucleon scattering cross sections. This shows that the ongoing and planned direct
detection experiments will play a vital role
in testing the bino-higgsino dark matter scenario (light blue points upper part on the $\sigma_{SI}- m_{\tilde{\chi}^{0}_1}$ plane).

\begin{table}[t!]\vspace{0.1cm}
\centering
\begin{tabular}{|p{2cm}|p{3cm}p{3cm}p{3cm}p{2.5cm}|}
\hline
\hline

                 	&	Point1	&	Point 2	&	Point 3	&	 Point4	\\
\hline

$m_0$	&$	          741	$&$	          1270	$&$	          793	$&$	          4270	$\\
$M_1$	&$	         -445	$&$	         -467	$&$	         -455	$&$	         -103	$\\
$M_2$	&$	          6.67\times 10^{+03}	$&$	          7010	 $&$	          6830	$&$	          1550	$\\
$M_3$	&$	          2.22\times 10^{+03}	$&$	          2340	 $&$	          2280	$&$	          516	$\\
$A_0$	&$	         -1990	$&$	         -2270	$&$	         -2110	$&$	         -7220	$\\
$\tan\beta$      	&$	15.32	$&$	19.15	$&$	15.49	$&$	 38.55	$\\
\hline		  		  		  		  	
$\mu$            	&$	115	$&$	101	$&$	257	$&$	2408	$\\

\hline		  		  		  		  	
$m_h$            	&$	125	$&$	125.4	$&$	125.2	$&$	125.2	 $\\
$m_H$            	&$	4059	$&$	4258	$&$	4154	$&$	3070	 $\\
$m_A$            	&$	4032	$&$	4230	$&$	4127	$&$	3050	 $\\
$m_{H^{\pm}}$    	&$	4060	$&$	4258	$&$	4155	$&$	3071	 $\\
		  		  		  		  	
\hline		  		  		  		  	
$m_{\tilde{\chi}^0_{1,2}}$	&$	         113,          122	$&$	         100,          107	$&$	         216,          267	$&$	          44,         1337	$\\

$m_{\tilde{\chi}^0_{3,4}}$	&$	         233,         5497	$&$	         244,         5784	$&$	         281,         5636	$&$	        2406,         2408	$\\

$m_{\tilde{\chi}^{\pm}_{1,2}}$	&$	         124,         5489	 $&$	         109,         5775	$&$	         275,         5629	 $&$	        1342,         2411	$\\

$m_{\tilde{g}}$  	&$	4633	$&$	4871	$&$	4737	$&$	1367	 $\\
		  		  		  		  	
\hline $m_{ \tilde{u}_{L,R}}$	&$	        5672,         3943	 $&$	        6037,         4258	$&$	        5801,         4036	 $&$	        4454,         4363	$\\
                 		  		  		  		  	
$m_{\tilde{t}_{1,2}}$	&$	        1660,         5056	$&$	        1810,         5339	$&$	        1689,         5167	 $&$	        1499,         2794	$\\
                 		  		  		  		  	
\hline $m_{ \tilde{d}_{L,R}}$	&$	        5673,         3946	 $&$	        6038,         4261	$&$	        5802,         4039	 $&$	        4455,         4366	$\\
                 		  		  		  		  	
$m_{ \tilde{b}_{1,2}}$	&$	        3806,         5118	$&$	        4041,         5406	$&$	        3892,         5231	 $&$	        2811,         3359	$\\
                 		  		  		  		  	
\hline		  		  		  		  	
$m_{\tilde{\nu}_{1}}$	&$	4220	$&$	4549	$&$	4319	$&$	 4379	$\\
                 		  		  		  		  	
$m_{\tilde{\nu}_{3}}$	&$	4212	$&$	4520	$&$	4310	$&$	 3829	$\\
                 		  		  		  		  	
\hline		  		  		  		  	
$m_{ \tilde{e}_{L,R}}$	&$	        4217,          686	$&$	        4545,         1252	$&$	        4316,          741	 $&$	        4376,         4267	$\\
                		  		  		  		  	
$m_{\tilde{\tau}_{1,2}}$	&$	         348,         4197	$&$	         871,         4505	$&$	         401,         4295	$&$	        3041,         3832	$\\
                		  		  		  		  	
\hline		  		  		  		  	
		  		  		  		  	
$\sigma_{SI}({\rm pb})$	&$	  1.76\times 10^{-09}	$&$	  1.25\times 10^{-09}	$&$	  5.70\times 10^{-09}	$&$	  9.20\times 10^{-14}	$\\

$\sigma_{SD}({\rm pb})$	&$	  1.47\times 10^{-04}	$&$	  1.29\times 10^{-04}	$&$	  1.35\times 10^{-04}	$&$	  2.44\times 10^{-09}	$\\

$\Omega_{CDM}h^{2}$	&$	  5.98\times 10^{-03}	$&$	  5.31\times 10^{-03}	$&$	  1.10\times 10^{-01}	$&$	  6.83\times 10^{+01}	$\\
\hline		  		  		  		  	
$\Delta_{EW}$	&$	  18.2	$&$	  12.8	$&$	  18.7	$&$	  1430	 $\\
		  		  		  		  	
$\Delta_{HS}$	&$	  137	$&$	  402	$&$	  169	$&$	  5820	 $\\

\hline
\hline
\end{tabular}
\caption{ Point 1 displays solution with  minimal value of $\Delta_{HS}$ . Point 2 represents minimal value of $\Delta_{EW}$
and $\Delta_{HS}$. Point 3 depict solutions corresponding minimal $\Delta_{EW}$ and
$\Delta_{HS}$ and best $\Omega_{CDM}h^{2}$ values. Point 4 displays Z-resonance channels.
}
\label{tab1}
\end{table}

It should be noted that IceCube currently
is sensitive only to relatively large neutralino masses and therefore does
not constrain the parameter space we have
considered. Likewise, while Super-K is sensitive in this region, the
bounds are not stringent enough to rule out anything. However,
from $\sigma_{SD}- m_{\tilde{\chi}^{0}_1}$ plane  we see that the future IceCube DeepCore
experiment will be able to constrain a significant region of the parameter space. In particular, it will test the bino-higgsino dark matter scenario.

Finally, we present a few benchmark points in Table 1 highlighting the phenomenologically interesting
features of  $Case~I$.
 All of these points are
consistent with neutralino LSP and the constraints
mentioned in Section~\ref{so10-model}. Point 1 displays a solution with  minimal value of $\Delta_{HS}$. Point 2 represents minimal values of $\Delta_{EW}$
and $\Delta_{HS}$. Point 3 depicts solutions corresponding to minimal $\Delta_{EW}$ and
$\Delta_{HS}$ and best $\Omega_{CDM}h^{2}$ values. Point 4 displays $Z$-resonance channels which is forbidden in the  universal gauginos ($M_{GUT}$) scenario due to the current gluino mass bound    $m_{\tilde{g}}>900$ GeV.  This solution is allowed in $Case~I$ because of the gaugino relation presented in Eq. (\ref{case1}). Although the fine tuning parameters are large for Point 4, it is still interesting to emphasize the existence of this solution.

\subsection{$ Case~II$ \label{modelI}}

\begin{figure}[th!]
\begin{center}
\includegraphics[width=7.6cm,height=6.cm]{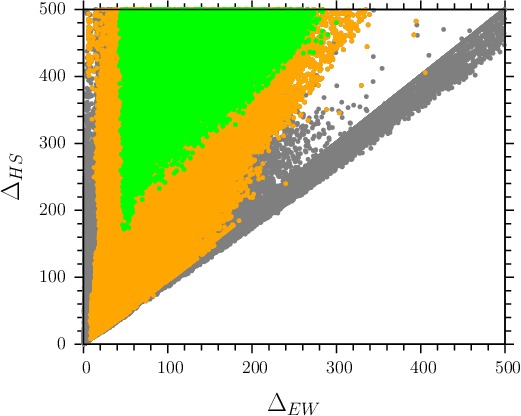}
\end{center}
\caption{Plots in  the $\Delta_{HS}- \Delta_{EW}$ planes for the $Case~II$. Gray points are consistent with REWSB and neutralino to be LSP. The orange points form a subset of the
gray ones and satisfy all the constraints described in Section \ref{so10-model}.  Green points belong to the subset of orange points
and satisfy the Higgs mass range $123\, {\rm GeV} \leq m_h \leq 127 \,{\rm GeV}$. }
\label{fig-d3}
\end{figure}

\begin{figure}[]
\begin{center}
\includegraphics[width=7.2cm,height=6.cm]{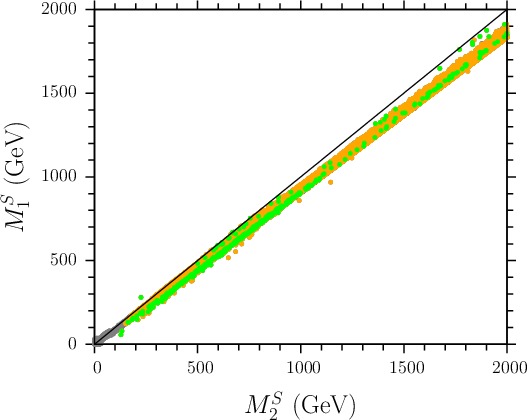}
\includegraphics[width=7.2cm,height=6.cm]{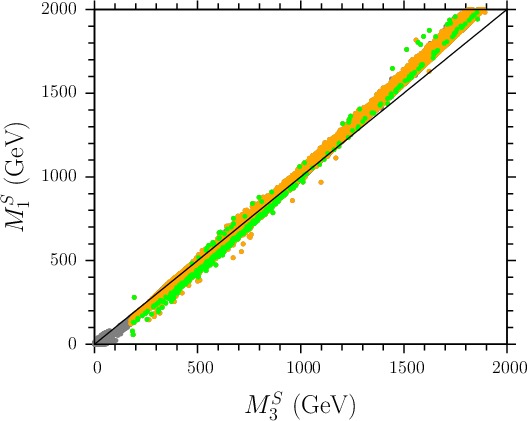}\vspace*{3mm} \\
\includegraphics[width=7.2cm,height=6.cm]{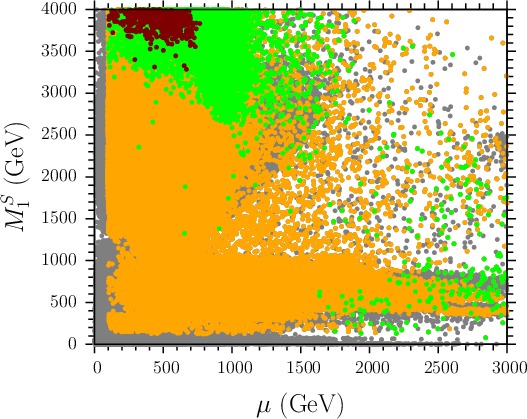}
\includegraphics[width=7.2cm,height=6.cm]{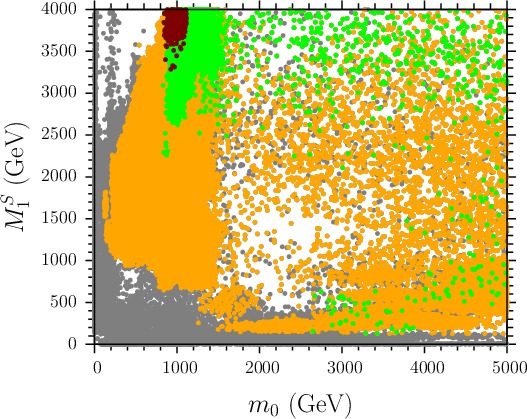}
\end{center}
\caption{Plots in $M^S_1 - M^S_2$, $M^S_3-M^S_1$, $M^S_1 - \mu$ and $m_0-M^S_1$ planes for the $Case~II$. Color coding
is the same as described in Figure \ref{fig-d3}.   In addition, we have used maroon color to denote a subset of the green points with $\Delta_{HS}<300$ and $\Delta_{EW}<150$.}
\label{fig-dd1}
\end{figure}

Next we consider  $Case~II$ for which the GUT scale gaugino masses satisfy  relation $M_1: M_2:M_3= -5:3:1$.
Figure \ref{fig-d3} shows our results in the $\Delta_{HS}-\Delta_{EW}$ planes for $Case~II$. Gray points are consistent with REWSB and LSP neutralino. Orange points form a subset of the gray ones and satisfy all the constraints described in Section \ref{so10-model}. Green points belong to the subset of orange points and satisfy the condition $123\, {\rm GeV} \leq m_h \leq 127 \,{\rm GeV}$.
The green points in the $\Delta_{HS}-\Delta_{EW}$ plane indicate that it is possible to have simultaneously $\Delta_{EW}\approx 50$ and
$\Delta_{HS}\approx 170$, which are larger compared to what we obtained for  $Case~I$. This increase in the values
of $\Delta_{HS}$  and $\Delta_{EW}$ can be understood from the results presented in Figure \ref{fig-dd1}.

Plots in $M_1^S-\mu$ and $m_0-M_1^S$ planes in Figure \ref{fig-dd1} show how the parameter space changes if a  solution of the little hierarchy problem is required. In this particular case the maroon color denotes a subset of the green points with $\Delta_{HS}<300$ and $\Delta_{EW}<150$. We see that the maroon points correspond to $M_1^S>3$ TeV and $m_0>1$ TeV. The $M_1^S-M_2^S$ and $M_3^S-M_1^S$ planes indicate the presence of approximate gaugino mass degeneracy at low scale. This happens because  in $Case ~II$, $M_1: M_2:M_3= -5:3:1$ at $M_{GUT}$  is almost the reverse of the relation at low scale, $M^S_1: M^S_2:M^S_3= 1:2:6$,  if we start with universal GUT scale gaugino mass relation.  To summarize  Figure \ref{fig-dd1},  the sparticle spectrum compatible with the little hierarchy problem in  $Case ~II$ is very heavy and will be difficult to observe at the LHC. A more detailed  discussion about the sparticle spectrum is presented below.


\begin{figure}[]
\begin{center}
\includegraphics[width=7.2cm,height=6.cm]{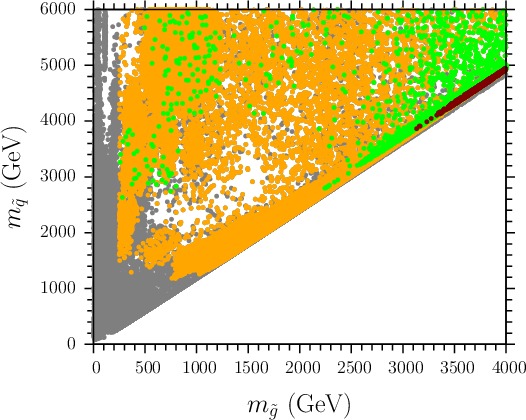}
\includegraphics[width=7.2cm,height=6.cm]{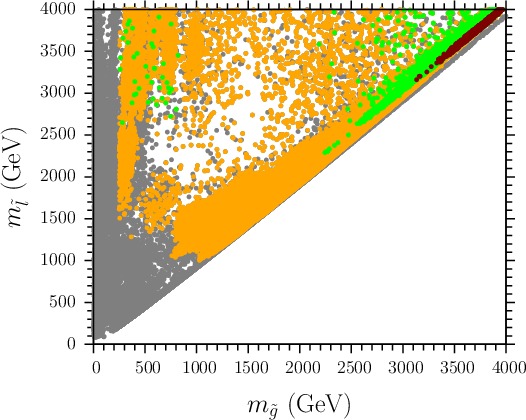}\vspace*{3mm}
\includegraphics[width=7.2cm,height=6.cm]{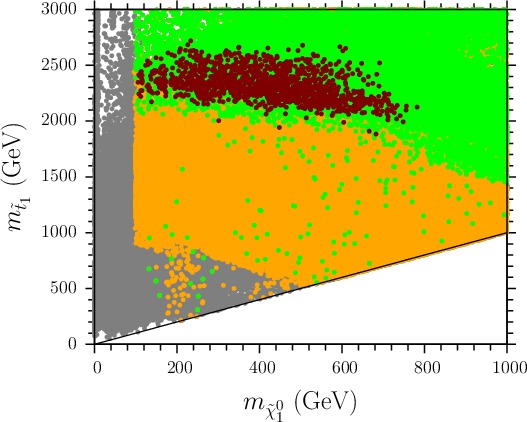}
\includegraphics[width=7.2cm,height=6.cm]{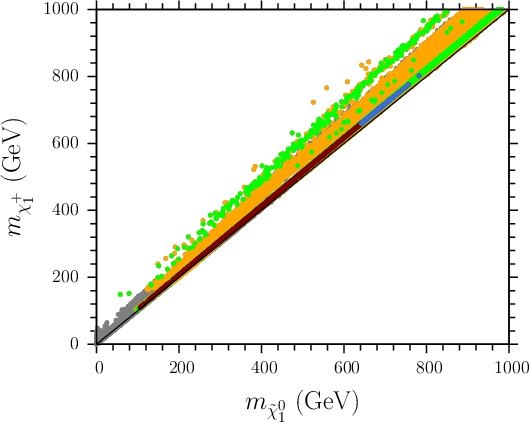}
\end{center}
\caption{Plots in $\Delta_{EW} - m_{h}$ and $\Delta_{HS} - m_{h}$ planes for the  $Case~II$. Color coding
is the same as described in Figure \ref{fig-dd1}.  In addition, we have used blue  color to denote a subset of the maroon points with $\Omega h^2<1$. }
\label{fig-d4}
\end{figure}


In Figure \ref{fig-d4}, we present the supersymmetric  particle spectrum  in the 
$m_{\tilde{q}}-m_{\tilde{g}}$,  $m_{\tilde{l}}-m_{\tilde{g}}$, $m_{\tilde{t}_{1}}-m_{\tilde{\chi}^{0}_1}$
and $m_{\tilde{\chi}^{+}_1}-m_{\tilde{\chi}^{0}_1}$ planes for  $Case~II$. Color coding
is the same as described in Figure \ref{fig-dd1}. In addition, we have used blue  color to denote a subset of the maroon points with $\Omega h^2<1$.  The $m_{\tilde{q}}-m_{\tilde{g}}$ panel shows that in $Case~II$, the  resolution of the little hierarchy problem with acceptable Higgs boson  mass  yields  gluino  and first two family squark masses heavier than 3 TeV.   These intervals are fully compatible with the current lower mass bounds obtained by the ATLAS and CMS collaborations. Figure \ref{fig-dd1} shows that the gauginos  are nearly degenerate with masses greater than 3 TeV. This  result explains why in the  $m_{\tilde{l}}-m_{\tilde{g}}$ plane we cannot have light sleptons compatible with the little hierarchy problem and, in general, the green points show that the sleptons in this model cannot be lighter than 2 TeV.

The lightest colored particle in this model is the stop quark. We can see from the  $m_{\tilde{t}_{1}}-m_{\tilde{\chi}^{0}_1}$ panel that the stop quarks cannot be lighter than 2 TeV, and the stops belong to the sparticle spectrum which ameliorate the little hierarchy problem. Moreover, if we relax the condition for natural supersymmetry, we can have neutralino-stop mass degeneracy at around 500 GeV, which is still compatible with the LHC data. To have a stop quark of around 500 GeV as the only light colored sparticle and all other colored sparticles above 2.5 TeV results from a combination of the RGE running and an interplay of parameters in the stop quark mass matrix \cite{Gogoladze:2011be}.

In the  $m_{\tilde{\chi}^{+}_1}-m_{\tilde{\chi}^{0}_1}$ plane we observe that the maroon points 
 lie near the unit line, which indicates that the
lightest neutralino  is  higgsino-like. This observation agrees with the results displayed in the $M_1^S - \mu$ plane  in Figure \ref{fig-dd1}. We can see that when $\mu$ is of order of 100 GeV, the bino has mass greater than 3 TeV.  This yields relatively low values for the  relic dark matter abundance unless the LSP neutralino mass is around 1 TeV \cite{Baer:2006te}. On the other hand
in the  $m_{\tilde{\chi}^{+}_1}-m_{\tilde{\chi}^{0}_1}$ plane we see some green points located  just above the unit line 
which, according to the results presented in the $M_1^S-M_2^S$ and $M_3^S-M_1^S$ planes in Figure \ref{fig-dd1}, are the solutions for neutralino  dark matter with bino-gluino \cite{Profumo:2004wk} and bino-wino \cite{Baer:2005jq} coannihilation channels.

Finally in Table~\ref{tab1} we present some benchmark points for $Case~II$.
Point 1 displays solution with the minimal value of $\Delta_{HS}$. Point 2 represents minimal values for $\Delta_{EW}$
and $\Delta_{HS}$. Point 3 depict solutions corresponding to minimal $\Delta_{EW}$ and
$\Delta_{HS}$ values and best $\Omega_{CDM}h^{2}$ values.

\begin{table}[t!]\vspace{0.1cm}
\centering
\begin{tabular}{|p{3cm}|p{3cm}p{3cm}p{3cm}|}
\hline
\hline

                 	&	Point 1	&	Point 2	&	Point 3	\\
\hline

$m_0$	&$	          1.31\times 10^{+03}	$&$	          1.31\times 10^{+03}	$&$	          2.37\times 10^{+03}	$\\
$M_1$	&$	         -1.11\times 10^{+04}	$&$	         -1.08\times 10^{+04}	$&$	         -8.66\times 10^{+03}	$\\
$M_2$	&$	          6.65\times 10^{+03}	$&$	          6.46\times 10^{+03}	$&$	          5.20\times 10^{+03}	$\\
$M_3$	&$	          2.22\times 10^{+03}	$&$	          2.15\times 10^{+03}	$&$	          1.73\times 10^{+03}	$\\
$A_0$	&$	         -3.82\times 10^{+03}	$&$	         -3.92\times 10^{+03}	$&$	         -4.72\times 10^{+03}	$\\
$\tan\beta$      	&$	30.94	$&$	35.15	$&$	44.38	$\\
\hline		  		  		  	
$\mu$            	&$	571	$&$	756	$&$	1498	$\\
$\Delta(g-2)_{\mu}$  	&$	  2.10\times 10^{-11}	$&$	  2.72\times 10^{-11}	$&$	  4.53\times 10^{-11}	$\\

\hline		  		  		  	
$m_h$            	&$	125	$&$	125.01	$&$	125.05	$\\
$m_H$            	&$	3868	$&$	3472	$&$	2426	$\\
$m_A$            	&$	3842	$&$	3449	$&$	2410	$\\
$m_{H^{\pm}}$    	&$	3869	$&$	3473	$&$	2428	$\\
		  		  		  	
\hline		  		  		  	
$m_{\tilde{\chi}^0_{1,2}}$	&$	         587,          587	$&$	         775,          775	$&$	        1517,         1518	$\\

$m_{\tilde{\chi}^0_{3,4}}$	&$	        5094,         5485	$&$	        4948,         5331	$&$	        3975,         4310	$\\

$m_{\tilde{\chi}^{\pm}_{1,2}}$	&$	         604,         5464	$&$	         795,         5311	$&$	        1543,         4290	$\\

$m_{\tilde{g}}$  	&$	4683	$&$	4556	$&$	3762	$\\
		  		  		  	
\hline $m_{ \tilde{u}_{L,R}}$	&$	        5809,         4923	$&$	        5659,         4800	$&$	        5052,         4438	$\\
                 		  		  		  	
$m_{\tilde{t}_{1,2}}$	&$	        2807,         4892	$&$	        2712,         4691	$&$	        2294,         3867	$\\
                 		  		  		  	
\hline $m_{ \tilde{d}_{L,R}}$	&$	        5810,         4324	$&$	        5660,         4222	$&$	        5052,         4051	$\\
                 		  		  		  	
$m_{ \tilde{b}_{1,2}}$	&$	        3722,         4929	$&$	        3451,         4727	$&$	        2892,         3895	$\\
                 		  		  		  	
\hline		  		  		  	
$m_{\tilde{\nu}_{1}}$	&$	4799	$&$	4673	$&$	4329	$\\
                 		  		  		  	
$m_{\tilde{\nu}_{3}}$	&$	4638	$&$	4462	$&$	3917	$\\
                 		  		  		  	
\hline		  		  		  	
$m_{ \tilde{e}_{L,R}}$	&$	        4797,         4268	$&$	        4671,         4158	$&$	        4327,         3966	$\\
                		  		  		  	
$m_{\tilde{\tau}_{1,2}}$	&$	        3859,         4627	$&$	        3625,         4451	$&$	        2943,         3906	$\\
                		  		  		  	
\hline		  		  		  	
		  		  		  	
$\sigma_{SI}({\rm pb})$	&$	  1.35\times 10^{-11}	$&$	  1.65\times 10^{-11}	$&$	  5.82\times 10^{-11}	$\\

$\sigma_{SD}({\rm pb})$	&$	  1.76\times 10^{-08}	$&$	  1.25\times 10^{-08}	$&$	  1.14\times 10^{-08}	$\\

$\Omega_{CDM}h^{2}$	&$	  6.01\times 10^{-02}	$&$	  1.06\times 10^{-01}	$&$	  3.39\times 10^{-01}	$\\
\hline		  		  		  	
$\Delta_{EW}$	&$	  8.47\times 10^{+01}	$&$	  1.38\times 10^{+02}	$&$	  5.40\times 10^{+02}	$\\
		  		  		  	
$\Delta_{HS}$	&$	  4.52\times 10^{+02}	$&$	  5.08\times 10^{+02}	$&$	  1.86\times 10^{+03}	$\\

\hline
\hline
\end{tabular}
\caption{ Point 1 displays solution with  minimal value of $\Delta_{HS}$ . Point 2 represents minimal value of $\Delta_{EW}$
and $\Delta_{HS}$ . Point 3 depict solutions corresponding minimal $\Delta_{EW}$ and
$\Delta_{HS}$ and best $\Omega_{CDM}h^{2}$ values.
}
\label{tab2}
\end{table}

\section{Conclusion \label{conclusions}}

We have attempted to ameliorate the little hierarchy problem and thereby implement natural supersymmetry within the framework of SO(10) grand unification. Through a judicious choice of SO(10) fields with designated vacuum expectation values, we can realize  $M_2/M_3 \approx 3$, which can yield natural supersymmetry. Here $M_2$ and $M_3$ denote the $SU(2)_W$ and $SU(3)_c$ gaugino masses at $M_{GUT}$.

We consider two distinct scenarios, namely $M_1:M_2:M_3=-1/5:3:1$ ($Case~I$), and $M_1:M_2:M_3=-5:3:1$ ($Case~II$). We explore the parameter space of these models which yield small fine-tuning measuring parameters (natural supersymmetry) at the electroweak scale ($\Delta_{EW}$) as well as at high scale ($\Delta_{HS}$). These two models yield more or less similar solutions to the  little hierarchy problem, however the light sparticle spectrum differs significantly. Depending on the ratio of the bino mass ($M_1$) to the other gaugino masses,   we can have in $Case~I$ stau leptons around 100  GeV, while in  $Case~II$ the stau sleptons lie in the TeV region.
In $Case~I$ we can have a bino-like neutralino as light as 90 GeV and gluino heavier than 2 TeV or so,  while in  $Case~II$ the gluino and bino are nearly degenerate in mass and the bino cannot  be lighter than a TeV or so. Having a relatively light neutralino with sizable bino-higgsino mixture in $Case~I$ makes the model testable at the direct dark matter search experiments. Finally, we present a few benchmark points in Tables 1 and 2 highlighting phenomenologically interesting
features of  $Case~I$ and $Case~II$. Relaxing the constraint from the little hierarchy problem, we find that the LSP neutralino can be a suitable dark matter candidate
 with bino-gluino and bino-wino  coannihilation channels.

\section*{Acknowledgments}
We thank   M. Adeel Ajaib and Shabbar Raza for valuable discussions.
This work is supported in part by the DOE Grant No. DE-FG02-12ER41808. This work used the Extreme Science
and Engineering Discovery Environment (XSEDE), which is supported by the National Science
Foundation grant number OCI-1053575.

\newpage

\thispagestyle{empty}



\begin{thebibliography}{99}

\bibitem{:2012gk}
  G.~Aad {\it et al.}  [ATLAS Collaboration],
  Phys.\ Lett.\ B {\bf 716}, 1 (2012).

\bibitem{:2012gu}
  S.~Chatrchyan {\it et al.}  [CMS Collaboration],
  Phys.\ Lett.\ B {\bf 716}, 30 (2012).

\bibitem{Carena:2002es}
  M.~S.~Carena and H.~E.~Haber,
  Prog.\ Part.\ Nucl.\ Phys.\  {\bf 50}, 63 (2003) and references therein.


\bibitem{Aad:2012fqa}
  G.~Aad {\it et al.}  [ATLAS Collaboration],
  Phys.\ Rev.\ D {\bf 87}, 012008 (2013).


\bibitem{Chatrchyan:2012jx}
  S.~Chatrchyan {\it et al.}  [CMS Collaboration],
  JHEP {\bf 1210}, 018 (2012).



\bibitem{Bhattacherjee:2013gr}
  B.~Bhattacherjee, J.~L.~Evans, M.~Ibe, S.~Matsumoto and T.~T.~Yanagida,
  arXiv:1301.2336 [hep-ph].








\bibitem{Ajaib:2012vc}
  M.~A.~Ajaib, I.~Gogoladze, F.~Nasir and Q.~Shafi,
  Phys.\ Lett.\ B {\bf 713}, 462 (2012).


\bibitem{Djouadi:2005gj}
For a review  see A.~Djouadi,
  Phys.\ Rept.\  {\bf 459}, 1 (2008) and reference therein.





  \bibitem{b5}
P.~H.~Chankowski, J.~R.~Ellis and S.~Pokorski,
Phys.\ Lett.\ B \textbf{423}, 327 (1998);
P.~H.~Chankowski, J.~R.~Ellis, M.~Olechowski and S.~Pokorski,
Nucl.\ Phys.\ B \textbf{544}, 39 (1999);
G.~L.~Kane and S.~F.~King, 
Phys.\ Lett.\ B \textbf{451}, 113 (1999);
G.~L.~Kane, J.~D.~Lykken, B.~D.~Nelson and L.~T.~Wang,
Phys.\ Lett.\ B \textbf{551}, 146 (2003).




\bibitem{mssm}
See  for instance  S.~P.~Martin,
  hep-ph/9709356 and references therein.


\bibitem{Baer:2012mv}
  H.~Baer, V.~Barger, P.~Huang, D.~Mickelson, A.~Mustafayev and X.~Tata,
  arXiv:1210.3019 [hep-ph].


\bibitem{Gogoladze:2012yf}
  I.~Gogoladze, F.~Nasir and Q.~Shafi,
  arXiv:1212.2593 [hep-ph].





\bibitem{Abe:2007kf}
  H.~Abe, T.~Kobayashi and Y.~Omura,
  Phys.\ Rev.\  D {\bf 76} (2007) 015002;
  I.~Gogoladze, M.~U.~Rehman and Q.~Shafi,
  Phys.\ Rev.\ D {\bf 80}, 105002 (2009).






\bibitem{Younkin:2012ui}
  D.~Horton, G.~G.~Ross,
  Nucl.\ Phys.\  {\bf B830 } (2010)  221;
  J.~E.~Younkin and S.~P.~Martin,
  Phys.\ Rev.\ D {\bf 85}, 055028 (2012);
  S.~Antusch, L.~Calibbi, V.~Maurer, M.~Monaco and M.~Spinrath,
  arXiv:1207.7236 [hep-ph].



\bibitem{Chamseddine:1982jx}
  A.~H.~Chamseddine, R.~L.~Arnowitt and P.~Nath,
  Phys.\ Rev.\ Lett.\  {\bf 49}, 970 (1982).
R.~Barbieri, S.~Ferrara, and C.~A. Savoy,
{ Phys. Lett. B} {\bf 119},  343 (1982);
 L.~J. Hall, J.~D. Lykken, and S.~Weinberg,
  {Phys. Rev. D} {\bf 27},  2359 (1983);
E.~Cremmer, P.~Fayet, and L.~Girardello,
 { Phys. Lett. B} {\bf 122},   41 (1983);
N.~Ohta,
{ Prog.
  Theor. Phys.} {\bf 70},  542 (1983).





\bibitem{Pati:1974yy}
  J.~C.~Pati and A.~Salam,
  Phys.\ Rev.\  D {\bf 10}, 275 (1974).


\bibitem{c-parity}
  T.~W.~B.~Kibble, G.~Lazarides and Q.~Shafi,
  Phys.\ Lett.\  B {\bf 113}, 237 (1982);
  T.~W.~B.~Kibble, G.~Lazarides and Q.~Shafi,
  Phys.\ Rev.\  D {\bf 26}, 435 (1982);
  R.~N.~Mohapatra and B.~Sakita,
  Phys.\ Rev.\  D {\bf 21}, 1062 (1980).





\bibitem{ISAJET}
  F.~E.~Paige, S.~D.~Protopopescu, H.~Baer and X.~Tata,
  hep-ph/0312045.




\bibitem{Martin:2007gf}
  S.~P.~Martin,
  Phys.\ Rev.\ D {\bf 75}, 115005 (2007);
  T.~J.~LeCompte and S.~P.~Martin,
  Phys.\ Rev.\ D {\bf 85}, 035023 (2012) and references therein.











\bibitem{Martin:2009ad}
 B.~Ananthanarayan, P.~N.~Pandita,
  Int.\ J.\ Mod.\ Phys.\  {\bf A22}, 3229-3259 (2007);
  S.~Bhattacharya, A.~Datta and B.~Mukhopadhyaya,
  JHEP {\bf 0710}, 080 (2007);
 S.~P.~Martin,
  Phys.\ Rev.\  {\bf D79}, 095019 (2009);
    U.~Chattopadhyay, D.~Das and D.~P.~Roy,
  Phys.\ Rev.\  D {\bf 79}, 095013 (2009);
  A.~Corsetti and P.~Nath,
  Phys.\ Rev.\  D {\bf 64}, 125010 (2001)
   and references therein.





\bibitem{Hill:1983xh}
   Q.~Shafi and C.~Wetterich,
  Phys.\ Rev.\ Lett.\  {\bf 52}, 875 (1984);
 C.~T.~Hill,
  Phys.\ Lett.\ B {\bf 135}, 47 (1984).





\bibitem{SO10GUT}
  H.~Georgi, Particles and Fields, {\it Proceedings of the APS
  Division of Particles and Fields}, ed.~C.~Carlson, p.~575 (1975).
  H.~Fritzsch and P.~Minkowski,
  Annals Phys.\  {\bf 93}, 193 (1975).



\bibitem{flipped} A. De Rujula, H. Georgi, and S.L. Glashow, {\it Phys.
Rev. Lett.} {\bf 45}, 413 (1980); H. Georgi, S.L. Glashow, and
M. Machacek, {\it Phys. Rev.} {\bf D23}, 783 (1981).



\bibitem{Aaltonen:2012ra}
  T.~Aaltonen {\it et al.}  [CDF and D0 Collaborations],
  Phys.\ Rev.\ D {\bf 86}, 092003 (2012).



\bibitem{Gogoladze:2012ii}
  I.~Gogoladze, Q.~Shafi and C.~S.~Un,
  JHEP {\bf 1207}, 055 (2012).


\bibitem{Leva}
J.L. Leva,  ACM Trans. Math. Softw. 18 (1992) 449-453;
J.L. Leva,  ACM Trans. Math. Softw. 18 (1992) 454-455.





\bibitem{Nakamura:2010zzi}
  K. Nakamura {\it et al.} [ Particle Data Group Collaboration ],
  J.\ Phys.\ G {\bf G37}, 075021 (2010).



















\bibitem{Baer:2002fv}
H.~Baer, C.~Balazs, and A.~Belyaev,
   { JHEP} {\bf 03} (2002) 042.






\bibitem{Aaij:2012hcp}
  R.~Aaij {\it et al.}  [LHCb Collaboration],
  CERN-PH-EP-2012-335.







\bibitem{Barberio:2008fa}
  E.~Barberio {\it et al.}  [Heavy Flavor Averaging Group],
  arXiv:0808.1297 [hep-ex].



\bibitem{Bennett:2006fi}
  G.~W.~Bennett {\it et al.}  [Muon G-2 Collaboration],
  Phys.\ Rev.\  D {\bf 73}, 072003 (2006).









\bibitem{Baer:2012cf}
  H.~Baer, V.~Barger, P.~Huang, D.~Mickelson, A.~Mustafayev and X.~Tata,
  arXiv:1212.2655 [hep-ph].




















\bibitem{Profumo:2004wk}
  S.~Profumo and C.~E.~Yaguna,
  Phys.\ Rev.\  D {\bf 69}, 115009 (2004);
  I.~Gogoladze, R.~Khalid and Q.~Shafi,
  Phys.\ Rev.\ D {\bf 79}, 115004 (2009)
  N.~Chen, D.~Feldman, Z.~Liu, P.~Nath and G.~Peim,
  Phys.\ Rev.\ D {\bf 83}, 035005 (2011)
  [arXiv:1011.1246 [hep-ph]].


\bibitem{Baer:2006te}
  N.~Arkani-Hamed, A.~Delgado and G.~F.~Giudice,
  Nucl.\ Phys.\ B {\bf 741}, 108 (2006);
  H.~Baer, A.~Mustafayev, E.~-K.~Park and X.~Tata,
  JCAP {\bf 0701}, 017 (2007).




\bibitem{Baer:2005jq}
  H.~Baer, T.~Krupovnickas, A.~Mustafayev, E.~K.~Park, S.~Profumo and X.~Tata,
  JHEP {\bf 0512}, 011 (2005);
  I.~Gogoladze, R.~Khalid and Q.~Shafi,
  Phys.\ Rev.\ D {\bf 80}, 095016 (2009).
  

\bibitem{Gogoladze:2011be}
  I.~Gogoladze, S.~Raza and Q.~Shafi,
  Phys.\ Lett.\ B {\bf 706}, 345 (2012).



\end{thebibliography}
\end{document}